\documentclass[a4paper,10pt]{article}
\usepackage{graphicx}

\begin{document}


\title{Extended regions of radio emission not associated with the AGN phenomenon as sources of acceleration of cosmic rays: The case of cD galaxies\footnote{Published in International Journal of Modern Physics: Conference Series, Vol. 8 (2012), 336, World Scientific Publishing Company}}

\author{ Gizani$^1$, Nectaria A. B.\\
$^1$ Physics Laboratory, School of Science and Technology, Hellenic Open University, \\ Patra, Greece \\}

\date{}
\maketitle

\paragraph{Abstract}
Diffuse, non-thermal extended emission not associated with the AGN phenomenon, found in many clusters of galaxies hosted by an AGN, are related to the acceleration of cosmic rays. 

In the current work we present preliminary evidence of absence of such formations in clusters of galaxies hosted by optically identified cD galaxies. 
Our subsample consists of three powerful low redshift radiogalaxies, centered in poor clusters of galaxies. We have searched for radio relics and (mini)halos which could be forming as a result of the confinement of cosmic rays by bubbles creayed by the AGN.  We report on the work in progress.
  \\


\section{Introduction}

Clusters of galaxies are the largest gravitationally bound systems and they are excellent cosmological probes. If they contain diffuse nonthermal components, then the latter can be used to trace past merger events between clusters, track the evolution of clusters in general and trace cosmic rays (CRs)
and magnetic fields (the constituents of intracluster medium).

Halos, relics and minihalos are extended low surface brightness regions of diffuse non-thermal radio emission of relativistic particles not directly associated with the active galactic nucleus hosting the cluster of galaxies or the other galaxies there. They are thought to be formed via (re)acceleration of various particles in weak shocks in the intracluster medium (ICM) or turbulence (see Ref.~\cite{ferrari}, and references therein). Based on numerical simulations ref.~\cite{pf} proposed a unified scheme of halos (minihalos) and relics: CR protons accumulate over the  life of the universe and are connected with non-equilibrium activities of clusters, while the primary CR electrons trace the current dynamical and non-equilibrium activity of forming structure (e.g. merger events). However ref.~\cite{uri} has suggested that all diffuse substructures arise from the same, homogeneous CR ion distribution and are manifestations of steady-state magnetic fields (flat spectrum halos) as well as irregular magnetic growth (relics and steep halos).  

Currently there are research works in the literature which connect the various features of radio emission from an AGN-host of cluster (e.g. references by Ensslin et al.). Radio ghosts result from former AGN radio lobes (e.g.~\cite{ensslin99}). That is they are formed when the central engine is not feeding the radio lobes via radio jets. The lobes are a storage of cosmic rays. When a merger event occurs, radio ghosts are revived (called radio phoenix) and their content is released in the ICM. This way radio halos~\cite{medina} and relics~\cite{gopal} are formed. In the X-rays we see cavities filled with radio emission or not (e.g.~\cite{macn}).

A cD galaxy is a very massive and extended brightest cluster galaxy located at the centre of a poor or dense cluster of galaxies. Most regular clusters, in the sense that they have reached stationary equilibrium, are usually hosted by cD galaxies. There are four proposed scenarios on the formation of a cD galaxy (e.g. ref~\cite{garijo}):  Accretion of gas which cools (cooling flow) forming stars in the cluster core, tidal stripping by cluster galaxies passing by the cluster centre, merging of galaxies during a cluster collapse and the most acceptable one merging of massive clusters. 

The study of diffuse non-thermal radio emission and cD galaxies is important as they can be used to trace the cosmological structure, i.e. the luminous matter large-scale distribution of the universe (see for example ref~\cite{west2} for the cD galaxies and ref~\cite{ensslinetal07} for the radio halos.

\section{Observations}

We search for halos/relics in a sample of 70 Abell clusters selected upon the radio to X-ray correlation from NVSS (NRAO VLA Sky Survey). Studying the clusters' diffuse components we can constrain particle acceleration processes and plasma magnetisation. Our scope is to probe their role in the production, acceleration and propagation of cosmic rays. In the current paper we present the study of three powerful AGN from our sample. These radiogalaxies (RGs) are situated at the center of dense galaxy cluster environments called 'Hercules A', '3C\,388' and '3C\,310'-cluster. We have used radio and X-ray data for all three radio sources: for Her A~\cite{gizanir}--\cite{gizanix}, for 3C388~\cite{rot}--\cite{leahyn}--\cite{kraft}, and for 3C310~\cite{van}--\cite{miller}  

\begin{figure}[pb]
\begin{center}
\centerline{\includegraphics*[width=8cm]{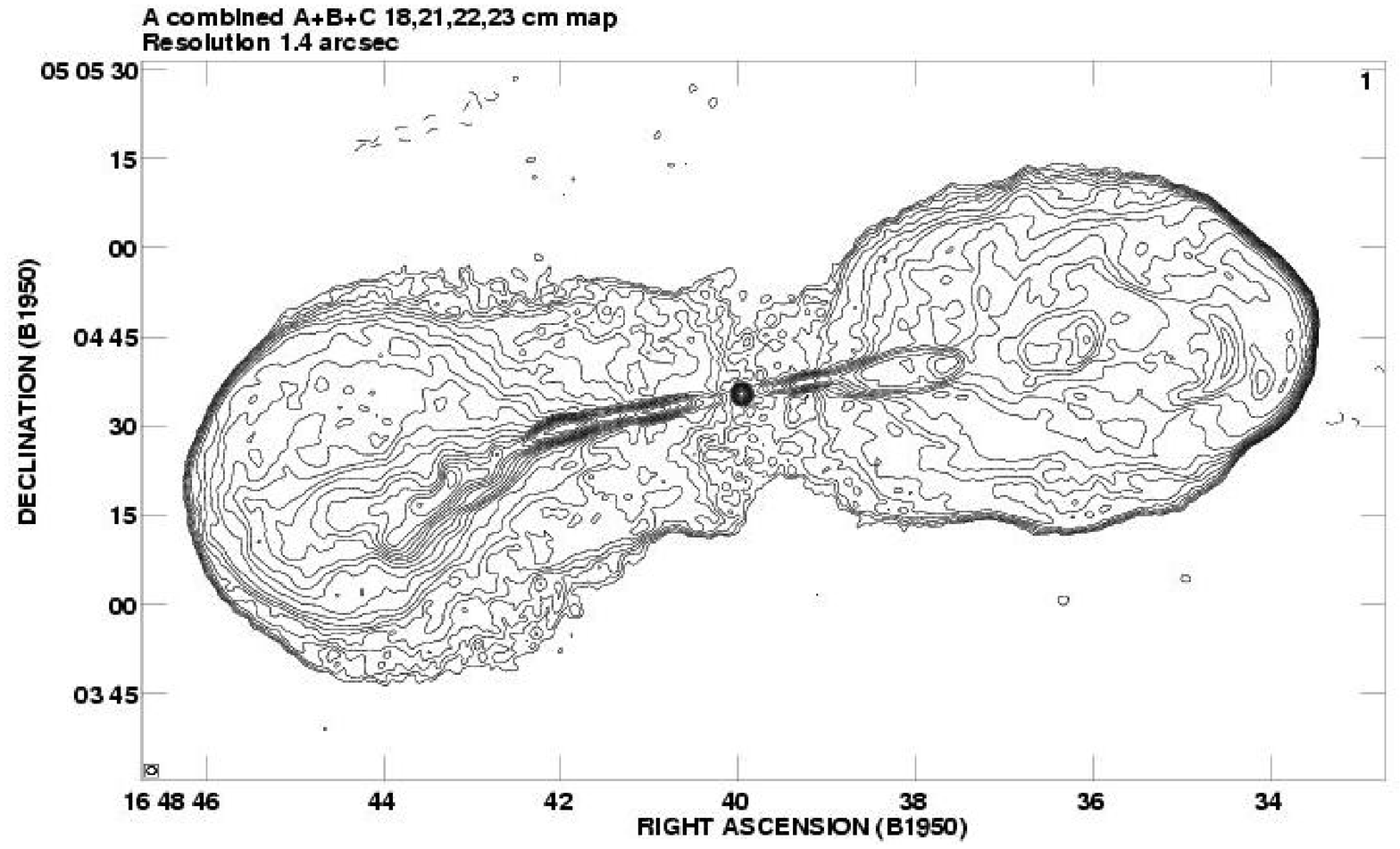}
\includegraphics*[width=8cm]{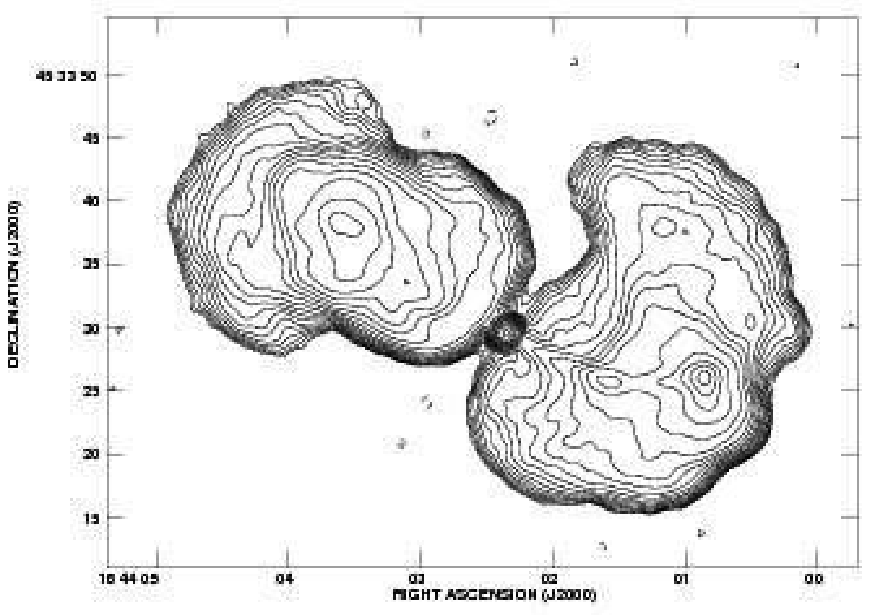}}
\end{center}
\vspace*{8pt}
\caption{The VLA map of Her A at a combined 20 cm frequency at 1.4 arcsec resolution (left from 12)
and of 3C388 at 21 cm at 1.3 arcsec resolution 
(right, from 15)
}
\label{f1}
\end{figure}

\begin{figure}[pb]
\begin{center}
\includegraphics*[width=5cm]{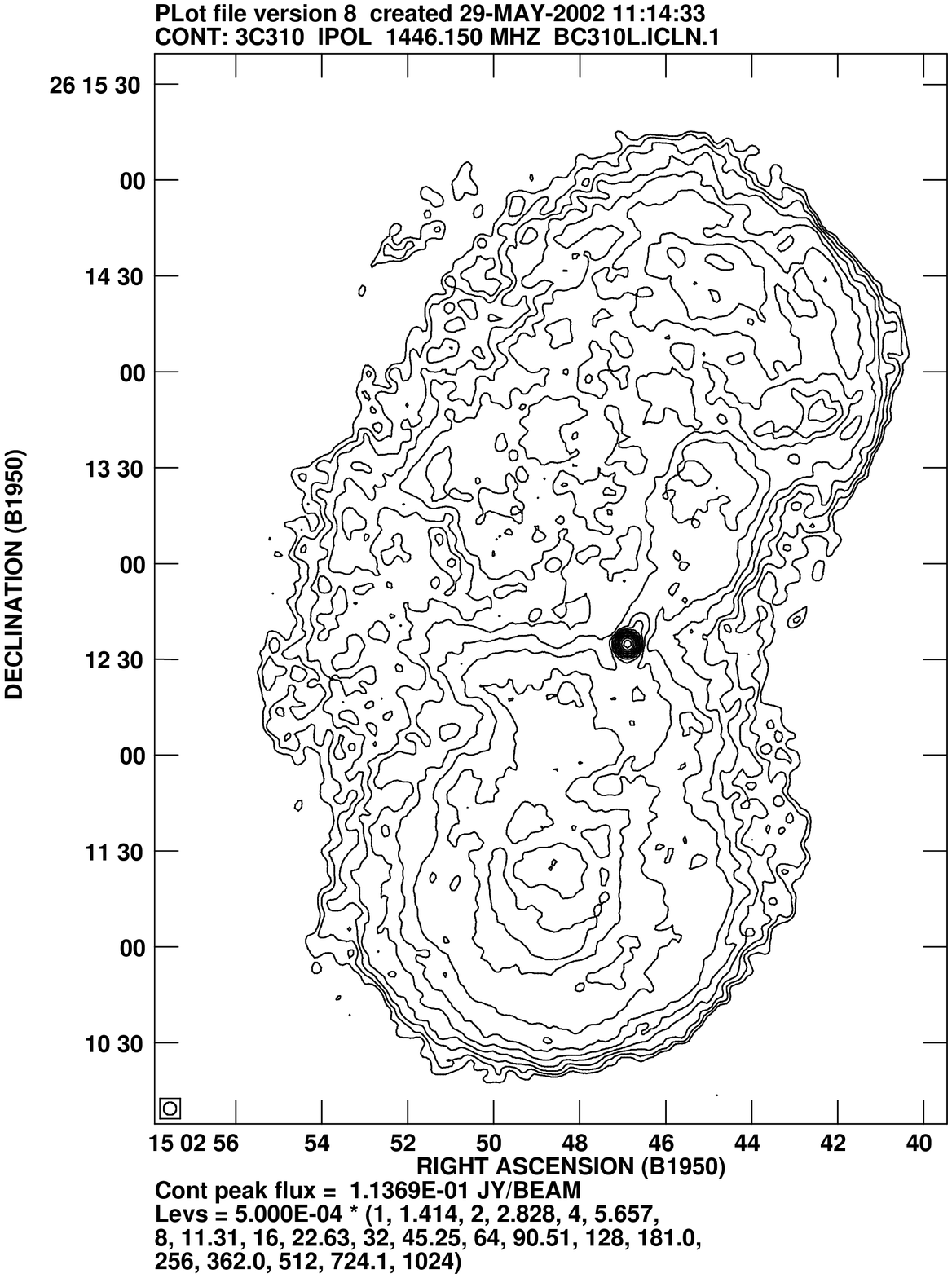}
\end{center}
\vspace*{8pt}
\caption{The VLA map of 3C310 at 21 cm at 4 arcsec resolution, adapted from 17.}
\label{f3}
\end{figure}

Figures~\ref{f1} (left from ~\cite{gizanir} and right from ~\cite{leahyn}) and~\ref{f3} (adapted from ~\cite{van}) show the kiloparsec scale radio VLA maps of the studied sources. None of the sources contain steep spectrum extended radio emission as a result of a merger event.

\section{Discussion}

In the current work, we have investigated the possibility powerful radio galaxies which are identified with cD galaxies in the optical to also present radio halos, relics and/or mini-halos. Our intuition came from Refs.~\cite{west2} and~\cite{west}, who suggested that cD galaxies - powerful radiosources are formed through anisotropic mergers. cD galaxies are now thought to be assembled via dissipationless major mergers (see for example ~\cite{carrasco}). Although substructures should not be expected in regular clusters containing cDs, radio observations have revealed their presence thus challenging the virialization and relaxation hypothesis. X-ray data shed light into the clusters' dynamical evolution.  

Hercules A, 3C\,388 and 3C\,310 were potential candidates for radio (mini-)halos/relics. They are all powerful radiosources identified with a cD galaxy residing in the centre of a cluster. In the X-rays, cavities have also been observed. However none of the studied RGs contain halos/relics which would be expected as a result of a merger event. Her A and 3C\,310 both contain flat spectrum radio arcs and rings ($\alpha \sim$0.8 and $\sim$ 1.0 respectively, $\alpha <$0) compared to the surrounding steep spectrum material ($\alpha \sim 1.5$), which are associated with the suppermassive black hole. The eastern lobe of 3C\,388 is embedded in a faint halo of steep spectrm and high polarization, interpreted to be a relic of previous epoch of jet activity (see for example~\cite{leahyn}). The ambiguity of a possible distortion of the gas by 3C\,388's radio lobes was clarified in the light of Chandra data~\cite{kraft}. The radiogalaxies are most probably relaxed doubles.

\end{document}